\documentclass[acmsmall,screen]{acmart}
\usepackage{graphicx}
\usepackage{textcomp}
\usepackage{adjustbox}
\usepackage{xcolor}
\usepackage{booktabs}
\usepackage{multirow}
\usepackage{adjustbox}
\usepackage{tikz}
\usetikzlibrary{arrows.meta, positioning}
\usepackage{float}
\usepackage{graphicx}  
\usepackage{caption}
\usepackage{hyperref}
\hypersetup{
    colorlinks=true,
    linkcolor=red,
    citecolor=blue,
    filecolor=blue,
    urlcolor=blue
}
\usepackage{subcaption}
\usepackage{algorithmic}
\usepackage[ruled, vlined, linesnumbered]{algorithm2e}
\usepackage{cleveref}  

\newcommand{\eg}{\hbox{\emph{e.g.}}\xspace}
\newcommand{\ie}{\hbox{\emph{i.e.}}\xspace}

\newcommand{\MyPara}[1]{\vspace{3pt}\noindent\textbf{\textit{#1}}\hspace{0.2em}}
\newcommand*{\tool}{HardRace\xspace}

\AtBeginDocument{%
  }

\setcopyright{acmlicensed}
\copyrightyear{}
\acmYear{}
\acmDOI{XXXXXXX.XXXXXXX}

\acmConference[Conference acronym 'XX]{Make sure to enter the correct
  conference title from your rights confirmation emai}{}{}
\acmISBN{}

\begin{document}

\title{HardRace: A Dynamic Data Race Monitor for Production Use}

\author{Xudong Sun}
\email{sxd@smail.nju.edu.cn}

\author{Zhuo Chen}
\email{ouhznehc@smail.nju.edu.cn}

\author{Jingyang Shi}
\email{jyshi@smail.nju.edu.cn}

\author{Yiyu Zhang}
\email{dz1933033@smail.nju.edu.cn}
\affiliation{%
  \institution{State Key Lab for Novel Software Technology, Nanjing University}
  \city{Nanjing}
  \country{China}
}

\author{Peng Di}
\email{dipeng.dp@antgroup.com}
\affiliation{
  \institution{Ant Group}
  \country{China}
}

\author{Jianhua Zhao}
\email{zhaojh@nju.edu.cn}

\author{Xuandong Li}
\email{lxd@nju.edu.cn}

\author{Zhiqiang Zuo}
\email{zqzuo@nju.edu.cn}
\orcid{0000-0001-7104-9918}
\authornote{Corresponding author.}
\affiliation{%
  \institution{State Key Lab for Novel Software Technology, Nanjing University}
  \city{Nanjing}
  \country{China}
}


\renewcommand{\shortauthors}{Sun et al.}

\begin{abstract}
Data races are critical issues in multithreaded program, leading to unpredictable, catastrophic and difficult-to-diagnose problems. Despite the extensive in-house testing, data races often escape to deployed software and manifest in production runs. Existing approaches  suffer from either prohibitively high runtime overhead or incomplete detection capability. In this paper, we introduce HardRace, a data race monitor to detect races on-the-fly while with sufficiently low runtime overhead and high detection capability. HardRace firstly employs sound static analysis to determine a minimal set of essential memory accesses relevant to data races. It then leverages hardware trace instruction, \ie, Intel PTWRITE, to selectively record only these memory accesses and thread synchronization events during execution with negligible runtime overhead. Given the tracing data, HardRace performs standard data race detection algorithms to timely report potential races occurred in production runs. The experimental evaluations show that HardRace outperforms state-of-the-art tools like ProRace and Kard in terms of both runtime overhead and detection capability -- HardRace can detect all kinds of data races in read-world applications while maintaining a negligible overhead, less than 2\% on average.


\end{abstract}





\maketitle

\section{Introduction}
In the post-Moore era, multithreaded software become prevalent; and concurrency errors become more and more common in multithreaded programs. Such errors cause critical issues such as program crashes \cite{study_asplos08}, security vulnerabilities \cite{Lazy_sosp17}, and incorrect computations \cite{hybrid_ppopp03}, leading to serious  real-world social and economic hazards, \eg, the Northeast blackout, and mismatched
Nasdaq Facebook share prices.

In spite of extensive in-house testing, data races often escape to deployed software and manifest in production runs \cite{tsan,empirical-lu-cocurrency}. 
This is because data races are highly sensitive to the execution states, including program inputs, thread interleavings, platform configurations, and other execution environments \cite{empirical-lu-cocurrency}. Such huge execution space can hardly be completely covered by testing. 
For the same reason, production-run data races are difficult to reproduce and fix offline \cite{clap-jeff}.
As a result, it is highly desirable to propose an online data race detector which is able to discover data races in production runs practically (with sufficiently low overhead) and effectively (with high accuracy).

\MyPara{Prior Work.} 
Over the past decades, extensive studies have been conducted, primarily classified as static and dynamic approaches. Static data race detectors, such as RacerD \cite{racerd} and CHESS \cite{chess},  analyze the code statically without executing it. It can report the potential races before deployment, thus being free of runtime overhead. However, due to the inherent limitation of over-approximation, the static approaches inevitably report many false positive warnings, severely affecting its practicability. 

On the other hand, dynamic data race detectors \cite{eraser,fasttrack,tsan}, monitor the program during execution to identify actual data races. 
Compared to static approaches \cite{racerd, chess} which suffer from high false positives, dynamic detectors have the advantage of high precision.
However, as they demand to collect and analyze massive execution data online, the program execution is  dramatically slowed down.  
For example, Google’s ThreadSanitizer (a.k.a., TSan) \cite{tsan}, the most mature and widely used tool in industry,  can result in an average 7-12x slowdown \cite{kard}. FastTrack also incurs the runtime overhead of a similar magnitude as reported \cite{fasttrack}.
Such high overheads severely inhibit the practical usage of these dynamic detectors -- they nowadays only work in debugging/testing mode but not production-run environment.

Recently, several attempts have been performed to lower the runtime overhead of dynamic race detectors \cite{prorace,kard,txrace,haccrg,racemob}. 
RaceMob \cite{racemob} adopts crowdsourcing, a user-level sampling mechanism to lower the runtime overhead of each individual user.
ProRace \cite{prorace} samples memory accesses using hardware PMU (in particular, Intel's Precise Event Based Sampling), thus achieving low runtime overhead. However, like all sampling approaches \cite{prorace,racez,pacer}, finding the right sampling rate is often challenging. Even worse, the detection accuracy is usually not guaranteed.
Kard\cite{kard} leverages Intel Memory Protection Keys (MPK) to achieve low detection overhead. However, due to the limitations of MPK, Kard can only detect a specific type of races, namely Inconsistent Lock Usage (ILU). Other common races remain undetectable.

\MyPara{Our Work.} 
This paper introduces \tool, a novel data race detector which leverages modern hardware tracing module (in particular, Intel Processor Trace instructions PTWRITE) to monitor data races in production runs, with sufficiently low overhead and high detection capability. 
However, naively employing hardware tracing for dynamic race detection faces two problems. 

First, naively tracing all the memory accesses via hardware still yields prohibitively high runtime overhead.
The reason is that data race detection requires tracking extensive runtime information, including memory accesses and thread synchronization events. 
To trace such dynamic information, a massive number of hardware tracing instructions  (\ie, \emph{ptwrite}) have to be instrumented and executed, leading to non-negligible overhead. 
Our empirical experiments show that the overhead of naive hardware tracing for data race detection reaches 19.8\% on average (see \S\ref{subsec:performance}). 

Second, naively recording the intensive memory access and thread synchronization information via hardware results in severe data loss, greatly diminishing detection capability. 
This is because the hardware generates traces much faster than memory can keep up. As a result, certain traces would be lost especially if the hardware trace instructions (\ie, \emph{PTWRITE}) is too dense \cite{jportal}. 
In our experiments, naive hardware tracing leads to frequent data loss, ~37\% on average, which significantly affects the detection capability (see \S\ref{subsec:effectiveness}). 

To tackle the above problems, \tool firstly employs effective static analysis to safely eliminate most memory accesses that are unlikely to be involved in data races. It then selectively instruments and traces only the remaining accesses via hardware.

\MyPara{Results.}
We implemented HardRace and evaluated it over a common set of benchmarks, including the widely used PARSEC/SPLASH-2x benchmark suite and a set of real-world applications with known data races \cite{pset}. 
The experimental results show that the runtime overhead of \tool is only around 1.6\% on average, which is significantly lower than that of the state-of-the-art approaches. 
Moreover, \tool can detect the data races in all the experimental subjects without any false negatives, whereas the existing dynamic detection tools like ProRace and Kard miss them a lot.   

This paper makes the following contributions:
\begin{itemize}
\item HardRace presents an effective data race detector with minimal overhead, that can be deployed to monitor production runs. 
\item HardRace firstly employs binary static analysis to safely prune away unnecessary memory accesses, and then leverage modern hardware tracing module (\ie, Intel PTWRITE) to realize selective tracing, achieving sufficiently low overhead and high detection precision. To the best of our knowledge, HardRace is the first to utilize Intel PTWRITE for precise and low-overhead data race detection.
\item The experimental evaluations show that HardRace outperforms state-of-the-art tools like ProRace and Kard in terms of both runtime overhead and detection capability -- HardRace can detect all kinds of data races in read-world applications while maintaining a negligible overhead compared to the existing solutions.
\end{itemize}

\MyPara{Outline.}
The rest of the paper is organized as follows. \S \ref{sec:background} gives the necessary background of hardware tracing and dynamic data race detection. 
\S \ref{sec:overview} provides the overview of HardRace. 
\S \ref{sec:online-trace} and \S \ref{sec:offline-analysis} describe the key components we proposed, followed by the implementation in \S \ref{sec:implement}.  
We present the empirical evaluations in \S \ref{sec:evaluate}. 
We talk about the related work in \S \ref{sec:related}. Finally, \S \ref{sec:conclude} concludes.

\section{Background}
\label{sec:background}
\subsection{Intel PTWRITE}

Intel PTWRITE is an extended hardware tracing feature, which is available at the commodity PCs with the 12th generation (Alder Lake) desktop processors. 
It provides the PTWRITE instruction to efficiently and flexibly record data values from registers or memory. If a register value of interest needs to be recorded, users can insert a PTWRITE instruction with the register as the operand. When the instruction is executed as the program runs, the hardware module writes the dynamic value of the register into a specific system buffer, which can be then read for usage. 

Intel PTWRITE is particularly advantageous for tracing multithreaded programs, where traditional software instrumentation often relies on expensive locking operations to determine the order of memory events across different threads. Such heavy intervention not only leads to a dramatic slowdown of the execution, but also significantly interferes with the original thread interleaving.
In contrast, Intel PTWRITE can efficiently and precisely record traces containing timestamp information (like TSC packets) thanks to dedicated hardware. When combined with timestamped thread-switching events recorded via OS tools (such as perf events), it becomes possible to associate the hardware trace packets (\ie, PTW packet) with each thread, thereby allowing for the reconstruction of a timeline of events across all threads.

Despite that Intel PTWRITE is of much lower overhead than traditional software techniques, it still encounters bottlenecks when recording the sheer volume of data. In particular, high-frequency recording scenarios can result in performance degradation and/or severe data loss \cite{herqules}. Thus, naively applying PTWRITE does not fully address the issue of multi-threaded recording. 
In this paper, we propose dedicated static analysis to eliminate the unnecessary tracing points, finally ensuring low-overhead and avoiding data loss.

\subsection{Data Race Detection}
Given the memory access traces, a race detection algorithm can be employed to determine if a race potentially happens --  two memory accesses may occur simultaneously.
Over the past years, various existing data race detection algorithms are proposed, including lockset \cite{eraser}, happens-before relation \cite{hb},  causally-precedes relation \cite{cp-relation}, weak-causally-precedes relation \cite{wcp-relation}, etc.  

The lockset algorithm, as its name suggests, focuses on lock/unlock accesses in multithreaded programs \cite{eraser}. Its primary principle is that shared variables must be protected by the same lock; otherwise, it is assumed that simultaneous access exists. However, not all multithreaded programs use locks to ensure logical non-concurrency. For instance, operations like signal and wait can also establish a temporal ordering. A sequence like \emph{<access A -> signal -> wait -> access A>} clearly does not constitute a data race. Therefore, the lockset algorithm is sometimes overly strict, leading to a significant number of false positives. 

The happens-before algorithm considers whether there is a sequential execution relationship between events across different threads (named as happens-before relationship) \cite{hb}. It avoids the false positives of lockset algorithm. However, its detection result is sensitive to the particular thread interleaving, meaning that even if one execution does not exhibit a race, it does not guarantee that other interleavings are also free of races. Thus, exploring a sufficient number of interleavings is necessary to reduce the risk of false negatives.
There are also other algorithms as the extension of happens-before relation, such as causally-precedes relation \cite{cp-relation} and weak-causally-precedes relation \cite{wcp-relation}. Basically, by relaxing the relation constrains, more races can be detected. 

In this paper, we propose an online data race monitor \tool, which focuses on reducing the runtime overhead without sacrificing detection capability. 
In brief, \tool first collects the necessary data access events at runtime, which are then fed into an existing detection algorithm to generate the report. 
For the sake of fair comparison, the offline analysis of \tool adopts the combination of happens-before and lockset algorithms which are used by the state-of-the-art dynamic race detectors \cite{tsan, fasttrack, prorace, kard}.
Note that the contribution of this paper is to propose a low-overhead and high-precision data access monitoring approach for multithreaded programs via hardware tracing, which is orthogonal to the race detection algorithm. The race detection algorithms can benefit from our lightweight monitoring; and \tool can also adopt any other detection algorithms in the offline analysis. 


\section{Overview}
\label{sec:overview}

\begin{figure}[ht]
    \centering
    \includegraphics[keepaspectratio, width=0.7\textwidth]{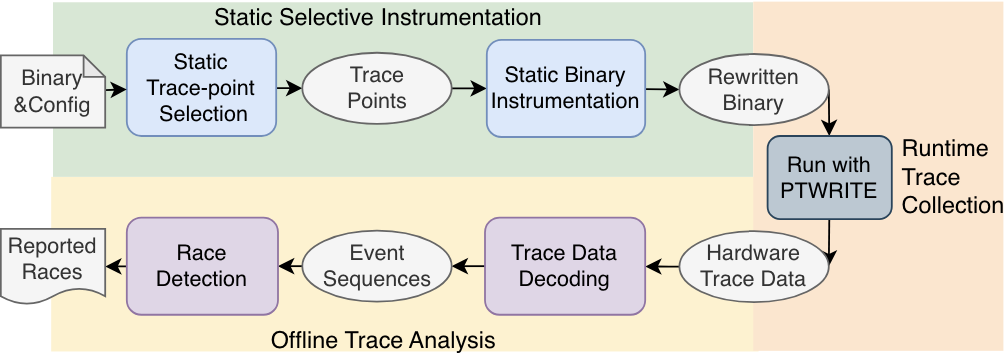}
   \caption{Workflow of HardRace}
   \label{fig@overview} 
\end{figure}

\tool is a data race monitor which can on-the-fly detect races happened in production runs. 
 \Cref{fig@overview} demonstrates the workflow of \tool, consisting of three main stages: static selective instrumentation, runtime trace collection, and  offline trace analysis. 

\MyPara{Static Selective Instrumentation.} The static selective instrumentation stage(see \S\ref{sec:online-trace}) is the core of our contribution. It is designed to statically identify a minimum set of memory accesses that are involved in data races and selectively instrument PTWRITE instructions to record them at runtime. 
It contains two key components: static trace-point selection (\S\ref{subsec:VSA}--\S\ref{subsec:recordElim}) and static binary instrumentation (\S\ref{subsec:SBI}). 
Specifically,  \tool takes the target binary file and some configuration settings as input. Static trace-point selection utilizes a series of sound static analysis algorithms to determine a set of memory accesses that must be not involved in data races and excludes them from the instrumentation points. Given a minimum set of memory access points to be recorded, the binary instrumentation module is responsible to insert PTWRITE instructions to record the identified data accesses and thread synchronization events. 

\MyPara{Runtime Trace Collection.} At this stage, the instrumented binary is executed on the CPUs with Intel PTWRITE supported in production runs. With the appropriate setting, hardware trace packets are continuously generated.  
We will give more implementation details in \S \ref{sec:implement}.

\MyPara{Offline Trace Analysis}. The offline analysis takes the hardware traces as input, and employs race detection algorithm to produce the final report. In particular, the tracked hardware traces are first decoded into per-thread memory access and synchronization event sequences (see \S\ref{subsec:decode}). These sequences are then passed to race detection algorithm for efficient data race detection (see \S\ref{subsec:detect}). 
The details about multithreaded-trace decoding and race detection will be elaborated in \S \ref{sec:offline-analysis} shortly. 

\begin{figure}[ht]
\centering
    \begin{minipage}{0.25\linewidth}
        \centering
        \begin{subfigure}{1\linewidth}
            \centering
            \includegraphics[width=\linewidth]{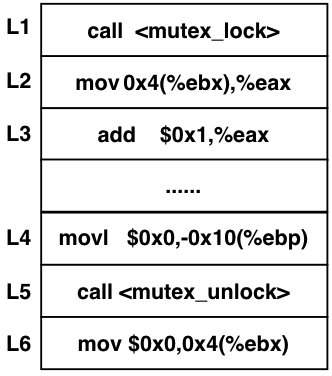}
            \caption{original binary}
            \label{fig@overview@examplea}
        \end{subfigure}\vspace{1em}
        \begin{subfigure}{0.7\linewidth}
        \centering
        \includegraphics[width=0.9\linewidth]{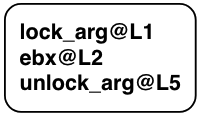}
        \caption{selected points}
        \label{fig@overview@exampleb}
        \end{subfigure}
    \end{minipage} \hspace{2em}
    \medskip
    \begin{minipage}{0.23\linewidth}
        \centering
        \raisebox{0mm}{
        \begin{subfigure}{1\linewidth}
        \includegraphics[width=\linewidth]{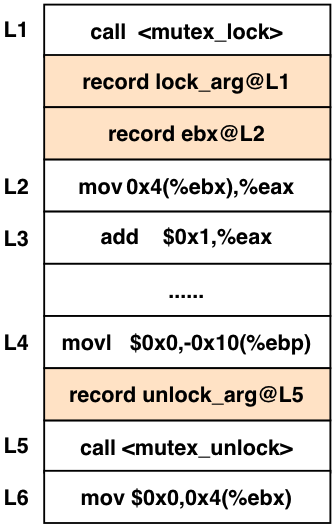}
        \caption{rewritten binary}
        \label{fig@overview@examplec}
        \end{subfigure}%
        }
    \end{minipage} \hspace{2em}
    \begin{minipage}{0.3\linewidth}
        \centering
        \begin{subfigure}{1\linewidth}
        \includegraphics[width=\linewidth]{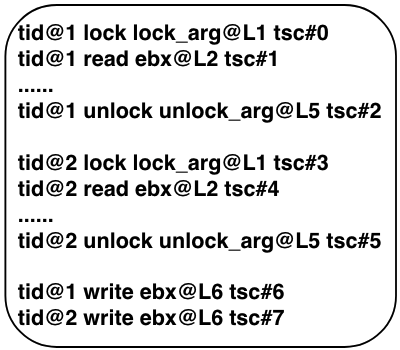}
        \caption{decoded output}
        \label{fig@overview@exampled}
        \end{subfigure}%
    \end{minipage} 
\caption{A toy example}
\label{fig@overview@example}
\end{figure}

\MyPara{Example.}
Let's use a toy example to illustrate the entire workflow of \tool. Suppose there is an assembly fragment in the original binary as shown in \Cref{fig@overview@examplea}. There are five instructions that involving memory accesses: \emph{L1, L2, L4, L5, and L6}. Naively, the values we need to record are \emph{lock\_arg@L1, ebx@L2, ebp@L4, unlock\_arg@L5 and ebx@L6}. 
After filtering by the selection module (\S \ref{sec:online-trace}), we can narrow down the trace points to \emph{lock\_arg@L1, ebx@L2, and unlock\_arg@L5}, as shown in \Cref{fig@overview@exampleb}. 
In this example, the two points \emph{ebp@L4 and ebx@L6} can be safely eliminated since \emph{ebp@L4} is a stack address so we can easily notice it is race-free. The value of register \emph{ebx@L6} can be statically deduced according to that of \emph{ebx@L2}.
During the static binary instrumentation, we insert three PTWRITE instructions into the binary, as illustrated in \Cref{fig@overview@examplec}. This rewritten binary is then executed during runtime trace collection, which generates trace data including thread information and PTWRITE packets. By decoding trace data, we get the sequence of memory accesses and synchronization events distinguished by threads, as shown in \Cref{fig@overview@exampled}. Here we assume there are two threads in the output, and each event records thread id, event type, access address and timestamp.
Having the decoded traces, we can directly employ the existing data race detection algorithm to obtain the final report. In this example, it is evident that the instructions at \emph{L2 and L6} access the same memory address in different threads \emph{tid@1 and tid@2}, and not be protected by same lock, which leading to a data race.

\section{Static Selective Instrumentation}
\label{sec:online-trace}
As mentioned above, the purpose of static selective instrumentation is to identify a minimum set of trace points including memory accesses and synchronization events. It filters out unnecessary trace points through a series of static analyses, thus reducing the runtime overhead. 
In the following, we will elaborate the static trace-point selection analysis in \S\ref{subsec:VSA} to \S\ref{subsec:recordElim}. 
Once having a subset of trace points, we then insert Intel PTWRITE instructions into the subject binary so as to record the necessary data at runtime. \S\ref{subsec:SBI} gives the detailed description about static binary instrumentation.

\begin{algorithm} 
    \KwIn{Binary $prog$}
    \KwOut{Rewritten binary $prog_{re}$}
    \caption{Static Selective Instrumentation}
    \label{alg:whole}
    \DontPrintSemicolon
    \BlankLine
    
    $ICFG \gets $ construct the interprocedural control flow graph of $prog$ \label{algo:icfg} \;
    $valueSetResult \gets $ \textsc{MultiVSA}($ICFG$) \label{algo:vsa} \textcolor{olive}{/*inter-procedural value set analysis for multithreaded programs (\S\ref{subsec:VSA})*/} \;
    $T_{shared} \gets $ \textsc{FindAllSharedAddresses}($valueSetResult$) \label{algo:shared} \textcolor{olive}{/*find all the race-relevant trace points involving shared variable addresses (\S\ref{subsec:VSA})*/} \;
    $T_{raceFree} \gets $ \textsc{MustRaceFree}($T_{shared}, valueSetResult$) \label{algo:must-race-free} \textcolor{olive}{/*identify the trace points which must not be relevant to data races  (\S\ref{subsec:MustRaceFree})*/} \;
    $T_{redundant} \gets $ \textsc{RedundantAnalysis}($T_{shared}-T_{raceFree}$) \label{algo:redundant} \textcolor{olive}{/*identify the redundant trace points whose values can be statically deduced (\S\ref{subsec:recordElim})*/} \;
    $T_{trace} \gets $ $T_{shared}-T_{raceFree}-T_{redundant}$ \label{algo:tracepoints} \;
    $prog_{re} \gets $ \textsc{Instrument}($prog, T_{trace}$) \label{algo:instrument} \textcolor{olive}{/*instrument PTWRITE instructions (\S\ref{subsec:SBI})*/} \;
\end{algorithm}

Algorithm \ref{alg:whole} provides a high-level description of how we progressively narrow down the set of trace points $T_{trace}$ that require instrumentation. At first, we construct the interprocedural control flow graph (ICFG) of the input program (Line \ref{algo:icfg}).
Next, we perform our inter-procedural value set analysis tailored for multithreaded programs to produce the binary-level alias results (Line \ref{algo:vsa}, see \S\ref{subsec:VSA}). 
Based on the value set results $valueSetResult$, we identify a set of trace points $T_{shared}$ that involve shared variable addresses, \ie, global and heap addresses (Line \ref{algo:shared}). Memory access points that only access stack addresses are filtered out at this stage, as they cannot cause data races.
Given the value set results $valueSetResult$ and all the potential trace points $T_{shared}$, a must race-free analysis is then performed to identify the set of trace points $T_{raceFree}$ that are impossible to cause races (Line \ref{algo:must-race-free}, see \S \ref{subsec:MustRaceFree}). 
Having the remaining trace points in $T_{shared}-T_{raceFree}$, we further identify the redundancies among them where the accessed addresses exhibit derivable relationships (Line \ref{algo:redundant}, see \S \ref{subsec:recordElim}). 
These redundant trace points are not necessary to be instrumented and traced at runtime since their values can be deduced during offline analysis.
With the three instruction sets identified, we compute $T_{trace} \gets T_{shared}-T_{raceFree}-T_{redundant}$ (Line \ref{algo:tracepoints}), which represents the set of trace points that actually need to be instrumented and tracked online. 
The static binary instrumentation module takes $T_{trace}$ as input  and produces the instrumented program $prog_{re}$ (Line \ref{algo:instrument}, see \S \ref{subsec:SBI}).

\subsection{Value-Set Analysis for Multithreaded Programs}
\label{subsec:VSA}
As mentioned above, in order to identify the set of trace points involving shared variable addresses $T_{shared}$, we need to determine from which memory region the operand (register) of an instruction originates, (\ie, stack region, heap region, or global region).  
We treat the registers stemming from a heap or global variable as shared addresses, which are likely involved in data races. The registers that are only relevant to stack addresses are excluded safely as they cannot cause races. 
To this end, we need a value-set analysis  to identify the relevant region (stack, heap, or global) for each register. 
Moreover, in the following must race-free analysis (\S \ref{subsec:MustRaceFree}), we also need the value-set analysis results to determine if two different registers may reference to the same memory location.

Value-set analysis \cite{wysinwyx} is a static binary analysis technique which over-approximates a set of values each register can take on at each program point. It can be considered as a binary-level alias analysis. 
Generally, value-set analysis involves two basic terms: abstract location and value set. 
An abstract location, or a-loc, is a variable-like entity, which can represent a register or an address in global, stack, and heap regions. For instance, for the instruction \emph{mov 0x4,\%eax}, both global address \emph{0x4} and the register \emph{\%eax} correspond to an a-loc.
A value-set represents a set of a-locs, and it is usually divided into three separate sets: stack, heap, and global. The value set of an a-loc is a collection of addresses and registers that can be accessed by referencing that a-loc. For the instruction \emph{mov 0x4,\%eax}, the value set of \emph{\%eax} would be $\langle global\mapsto\{0x4\}, stack\mapsto\{\}, heap\mapsto\{\} \rangle$, meaning that the  register \emph{\%eax}  holds the value \emph{0x4} from the global memory region after the instruction executes.

\begin{algorithm}
\SetKwData{Left}{left}\SetKwData{This}{this}\SetKwData{Up}{up}
\SetKwFunction{VSA}{VSA}
\SetKwProg{fn}{Function}{}{}
    \caption{Value-set analysis for multithreaded programs} \label{alg:VSA}
    \BlankLine
    \KwData{\emph{localValueSet[i][x]}, the value set of a-loc \emph{x} for instruction \emph{i} where \emph{x} is a register or stack a-loc; \emph{sharedValueSet[y]}, the value set of a-loc \emph{y} which is a global or heap a-loc}
    \BlankLine
    
    \emph{localValueSet} $\gets \emptyset$ \label{algo:vsa-localinit} \\
    \emph{sharedValueSet} $\gets \emptyset$ \label{algo:vsa-sharedinit} \\
    $worklist \gets$ put all the entry instructions into worklist \\
    \While{worklist $\neq \emptyset$}{
        $i \gets worklist.pop()$ \label{algo:vsa-pop} \\
        \emph{oldLocalValueSet[i]} $\gets$ \emph{localValueSet[i]} \\
        transfer(\emph{i, localValueSet, sharedValueSet}) \label{algo:vsa-transfer} \\
        propagate($i, succs(i)$) \label{algo:vsa-propagate} \\
        \If{oldLocalValueSet[i] $\neq$ localValueSet[i]}{
            worklist.push($succs(i)$)\\
        }
    }
\BlankLine
\fn{transfer(i, localValueSet, sharedValueSet)}{
    \uIf{i is mov}{
        \emph{src\_alocs} $\gets getAliasAlocs(\emph{i},\emph{src\_op}, \emph{localValueSet}, \emph{sharedValueSet})$\\
        \emph{dst\_alocs} $\gets getAliasAlocs(\emph{i},\emph{dst\_op}, \emph{localValueSet}, \emph{sharedValueSet})$\\
        \label{line:opHandle13}
        \ForEach{dst\_aloc $\in$ dst\_alocs}{
            \uIf{dst\_aloc is register or stack a-loc}{
                \emph{localValueSet[i][dst\_aloc]}.union(src\_alocs)
            }\Else{
                \textcolor{olive}{/*dst\_aloc is global or heap.*/}\\
                \emph{sharedValueSet[dst\_aloc]}.union(src\_alocs)    
            }
        }
    }\ElseIf{...}{
        ... \textcolor{olive}{/*The propagation of other instructions is omitted here.*/} 
    }
}
\end{algorithm}

Although multiple value-set analyses have been proposed \cite{wysinwyx,selectivetaint,vsa-refine}, none of them supports multithreaded programs well.
In other words, existing value-set analysis cannot ensure the soundness for multithreaded programs. 
To be specific, the existing value-set analysis maintains a value-set for a register at each instruction of a single thread. It does not consider the effects of shared (heap or global) accesses by multiple threads. Therefore, for a given register of an instruction, its value-set may not be a safe over-approximation of the actual values at runtime. 
As a consequence, we may erroneously exclude certain trace-points (registers) which are actually related to shared accesses, leading to loss of detection capability.
To guarantee soundness of value-set analysis, one way is to extend the control-flow graph by adding all the possible edges because of interleavings. Unfortunately, such approach could cause the control-flow graph to be amplified dramatically considering the large number of interleavings, thus leading to poor analysis scalability. 

In this paper, we devise a dedicated value-set analysis ensuring both soundness and scalability. Similar to the existing analysis, we record the value set of a register at each instruction along the traditional control-flow graph. However, for global and heap locations, we maintain a shared summary across the entire program. In this way, the value-set of shared accesses can be guaranteed to be an over-approximation of the actual values. thus ensuring soundness. Moreover, the analysis can scale well to large programs, since the analysis is performed along the original control-flow graph and the value-set relevant to global and heap locations are flow-insensitive.

Algorithm \cref{alg:VSA} shows the algorithm in details. 
At the beginning, \emph{localValueSet} and \emph{sharedValueSet}  are initialized as empty (Lines \ref{algo:vsa-localinit}-\ref{algo:vsa-sharedinit}), which represent the value-set results for stack and shared (global and heap) locations, respectively. 
The worklist algorithm processes each instruction, starting from the entry instruction of the program (Line \ref{algo:vsa-pop}). 
The transfer function is performed to update \emph{localValueSet[i]} and \emph{sharedValueSet} (Line \ref{algo:vsa-transfer}), followed by propagating the value-set to all the successors (Line \ref{algo:vsa-propagate}), until all \emph{localValueSet} entries remain unchanged. Since the value-set updates in the transfer function are performed using \emph{union} operation, the algorithm is guaranteed to converge and terminate.
In the transfer function, we update \emph{localValueSet} and \emph{sharedValueSet} according to the specific semantics of each type of instruction. Here, we only give the transfer logic of \emph{mov} instruction due to space limit.
Generally, it first retrieve the value-sets (\ie, \emph{localValueSet} and \emph{sharedValueSet}) to acquire the corresponding alias a-locs of the source and destination operands of instruction \emph{i}. 
Then, based on the type of destination a-loc, it decides to update \emph{localValueSet} or \emph{sharedValueSet} accordingly. 

Having the value-set results, we can obtain the set of trace points (\ie, registers) $T_{shared}$, which are relevant to shared (heap or global) memory regions.
In particular, given a register of an instruction, if its value-set contains any heap or global a-locs, we include it into $T_{shared}$. Otherwise, if the value-set of a register only contains stack addresses, we can safely exclude it from tracing.
Specifically, for an instruction \emph{i = mov 0x18(\%eax), \%edx}, we check \emph{localValueSet[i][eax]}. If it contains a-locs belonging to global or heap regions, then \emph{eax} could represent the address of a shared variable. We thus put $eax@i$ into $T_{shared}$. Otherwise, $eax@i$ can be safely eliminated.

\subsection{Must Race-free Analysis}
\label{subsec:MustRaceFree}

In this section, we would like to employ static analysis to further prune away the trace points irrelevant to data races. As is well known, data races occur only if three conditions are satisfied: 1) two memory accesses target the same address; 2) they are accessed concurrently; 3) at least one of them is a write operation.
Based on this, we propose a static must race-free analysis. In brief, for each memory trace-point in $T_{shared}$, we statically check if at least one of the above three conditions must be violated. If so, we consider it as race-free access. We can thus safely avoid it from tracing. 



\begin{algorithm*}[ht] \SetKwData{Left}{left}\SetKwData{This}{this}\SetKwData{Up}{up}\SetKwProg{fn}{Function}{}{}
    \KwIn{a set of shared trace-points $T_{shared}$}
    \KwOut{a set of trace-points that must be irrelevant to races $T_{raceFree}$}
    \caption{Must Race-free Analysis}
    \label{alg:raceFree}
    \BlankLine
    $T_{raceFree} \gets  \emptyset$, $T_{mayRace} \gets \emptyset$ \\
    \ForEach{x $\in T_{shared}$}{
        $flag_{raceFree} \gets true$\\
        \If{$x \notin T_{mayRace}$}{  
            \ForEach{$y \in T_{shared}$}{
                \uIf{$notAlias(x,y)$ or $notConcurrent(x,y)$ or $notWrite(x,y)$}{ \label{algo:racefree-check}
                    \textbf{continue}
                }\Else{
                    $flag_{raceFree} \gets false $\\
                    $T_{mayRace}.push(y)$\\
                    \textbf{break}
                }
            }
            \uIf{$flag_{raceFree}$}{
                $T_{raceFree}.push(x)$
            }\Else{
                $T_{mayRace}.push(x)$
            }
        }
    }
\BlankLine
    \fn{notConcurrent(x, y)}{\label{algo:notcon-begin}
    \If{$isOwned(x) \ or\ isOwned(y)$}{
        \textcolor{olive}{/*$isOwned$ means a heap object is allocated intra-procedurally and does not escape.*/}\\
        \Return true
    }
    \If{$getLockSet(x) \cap getLockSet(y) \neq \emptyset$}{
        \Return true
    }
    \Return false \label{algo:notcon-end} 
}
\BlankLine
\fn{IsOwned(x)}{\label{algo:own-begin}
    \emph{i} $\gets$ the instruction of $x$\\
    \emph{alocs} $\gets$ \emph{localValueSet[i][x]}\\
    \ForEach{aloc $\in$ alocs}{
        \If{aloc is heap and is allocated within the function of \emph{x}}{
            \ForEach{arg\_aloc: the argument of callsites within the function of \emph{x}}{
                \If{aloc $\in$ localValueSet[i][arg\_aloc]}{
                    \Return false \textcolor{olive}{/*escape to other function, not intra-procedural*/}\\
                }
            }
            \If{aloc $\in$ sharedValueSet}{
                \Return false \textcolor{olive}{/*escape to memory*/}\\
            }
        }    
        \Return false
    }
    \Return true \label{algo:own-end}
}

\end{algorithm*}


The core logic of the algorithm is shown in \Cref{alg:raceFree}. We take the set of trace-points $T_{shared}$ involving global/heap memory as input and iterate over all pairs of registers $x, y \in T_{shared}$. 
At Line \ref{algo:racefree-check}, we check if any one of the three conditions are violated. 
A trace point $x$ is considered to belong to $T_{raceFree}$ if and only if it does not form a data race with any possible point $y$. 
The rest trace points of $T_{shared}$ would be treated as $T_{mayRace}$. 

In the following, we elaborate how to check the three conditions. 
\emph{notWrite} can be checked easily by simply determining whether \emph{x or y} is a \emph{write} operation. 
The determination of \emph{notAlias(x, y)} is done on the basis of value-set analysis (\S \ref{subsec:VSA}).
The result of value-set analysis provides a set of potential values for a specific register at each program point, which essentially identifies the possible may-alias relationships between the register and different a-locs (\ie, other registers and addresses). 
As long as the intersection of the value-sets for \emph{x} and \emph{y} is empty, it can be concluded that \emph{x} and \emph{y} access different addresses, \ie, \emph{notAlias(x, y)} returns true.
To determine whether the concurrent access condition is met, the core logic of the \emph{notConcurrent} function is shown in Lines \ref{algo:notcon-begin}-\ref{algo:notcon-end}. 
Basically, it follows the logic of lockset algorithm.
If  $x$ or $y$ is an allocated heap object intra-procedurally and does not escape, we can reach that $x$ and $y$ cannot be executed concurrently. 
The detailed logic for determining \emph{isOwned} is shown as Lines \ref{algo:own-begin}-\ref{algo:own-end}.
More importantly, it also considers the accesses within critical sections enclosed by locks and unlocks. The $getLockSet$ function analyzes and returns the lock variables involved. If the intersection of $getLockSet(x)$ and $getLockSet(y)$ is non-empty, then $x$ and $y$ are protected by the same lock. Thus, true is returned.

\subsection{Redundant Register Elimination}
\label{subsec:recordElim}

After the trace-point elimination discussed above, the set $T_{shared}-T_{raceFree}$ may still have potential for further reduction. The rationale is that the addresses accessed in two instructions may have a static relationship. In other words, the value of a register in one instruction can be statically derived from that of another in other instruction. In such cases, we only need to record the deriving register. The values of subsequent registers can be deduced offline. 


\begin{figure}[ht]
\centering
\resizebox{0.75\textwidth}{!}{
\begin{tikzpicture}[
    every node/.style={draw, rectangle, minimum width=2cm, minimum height=1cm, text centered},
    block/.style={draw, rectangle, minimum width=3cm, minimum height=1cm, text centered, font=\ttfamily, align=center},
    arrow/.style={draw, -{Stealth[scale=1.2]}, thick}
]

\node (B1) [block] {
  \texttt{L1: mov 0x8(\%ebp),\%eax} \\
  \textbf{L2: mov 0x4(\%eax),\%eax} \\
  \texttt{L3: test \%eax,\%eax}\\
  \texttt{L4: je L5}
};

\node (B2) [block, below left =of B1] {
  \texttt{L5: mov ecx, edx} \\
  \texttt{L6: sub ecx, 1}
};

\node (B3) [block, below right=of B1] {
  \texttt{L7: mov -0xc(\%ebp),\%eax} \\
  \textbf{L8: mov \%edx,0x8(\%eax)} \\
  \texttt{\ldots}\\
  \textbf{L9: mov \%edx,0xc(\%eax)} \\
  \texttt{L10: jmp L11}
};

\node (B4) [block, below=of B2] {
  \texttt{L11: mov 0x8(\%ebp),\%eax} \\
  \textbf{L12: mov 0xc(\%eax),\%eax}\\
  \texttt{\ldots}
};

\draw [arrow] (B1) -- (B2);
\draw [arrow] (B1) -- (B3);
\draw [arrow] (B2) -- (B4);
\draw [arrow] (B3) -- (B4);

\end{tikzpicture}
}
\caption{A toy example for redundant register elimination}
\label{fig@reg_elim}
\end{figure}
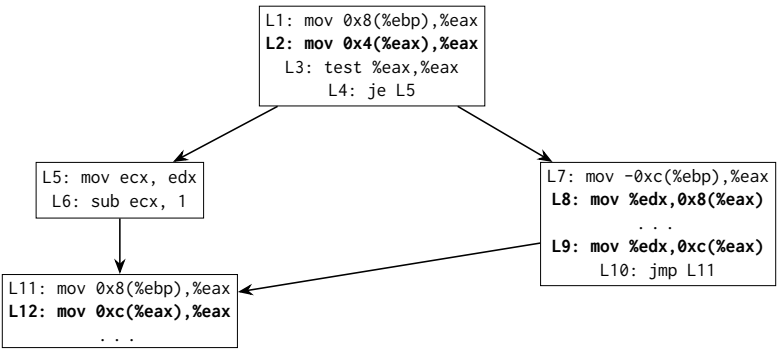

For example, \Cref{fig@reg_elim} has four basic blocks, and the memory instructions that need to be tracked are highlighted in bold (\ie, L2, L8, L9, and L12). 
For L8 and L9, there are two memory accesses, \emph{0x8(\%eax)@L8} and \emph{0xc(\%eax)@L9}.
Naively, we need to instrument and trace the value of \emph{eax} at both L8 and L9. 
In fact, \emph{eax} at both L8 and L9 has the same value, which equals to \emph{-0xc(\%ebp)} where \emph{ebp} is a local address.
Thus, it is sufficient to record the value of \emph{\%eax} only at \emph{L8}, rather than both.
Beyond the elimination within a basic block, we also consider the situations across basic blocks. For instance, the \emph{eax} at \emph{L2} and \emph{L12} are also identical no matter which branch is taken, which equals to \emph{0x8(\%ebp)@L1}. Therefore, only one \emph{eax} needs to be traced.


Technically, given a memory access register, we perform an intra-procedural backward symbolic propagation from it until the entry of function. If the symbolic expressions of two registers are identical or have statically fixed relation, then we only keep one as the trace-point. The value of another will be statically deduced.

\subsection{Static Binary Instrumentation}
\label{subsec:SBI}
Based on \S\ref{subsec:VSA} to \S\ref{subsec:recordElim}, we identified a minimal set of trace points to be recorded at runtime. 
Here,  we exploit hardware tracing module to achieve low runtime overhead. 
In particular, we do this by inserting a PTWRITE instruction with the operand being the specific register. 
Taking \Cref{fig@overview@example} as an example, for the instruction \emph{<mov 0x4(\%ebx), \%eax>} at location L2, the value to be recorded is \emph{ebx@L2}, so a \emph{<ptwrite \%ebx>} instruction is inserted right before L2.
Each register may be instrumented multiple times at different locations. Therefore, we need to distinguish which instruction each PTWRITE packet corresponds to.  
To this end, during instrumentation, we manually maintain the mapping between each PTWRITE instruction and the instruction to be traced. This allows us to further determine the exact register corresponding to the PTWRITE packet. 
For redundant register elimination in \S\ref{subsec:recordElim}, we maintain the arithmetic relationship between two registers. In the decoding phase, the eliminated memory accesses are reconstructed based on the recorded register value. 

\section{Offline Trace Analysis}
\label{sec:offline-analysis}
\subsection{Hardware Trace Decoding}
\label{subsec:decode}
The hardware traces generated by Intel PTWRITE is stored in a compact packet format. 
Before analyzing them, we need to firstly decode these packets into the memory read/write events and thread synchronization events required by the race detection algorithm on a per-thread basis. 

\begin{figure}[ht]
\centering
    \begin{minipage}{0.20\linewidth}
        \centering
        \begin{subfigure}{1\linewidth}
            \centering
            \includegraphics[width=0.75\linewidth]{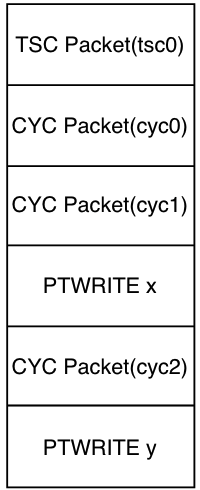}
            \caption{hardware packets}
            \label{fig@decode_a}
        \end{subfigure}
    \end{minipage} \hspace{2em}
    \medskip
    \begin{minipage}{0.20\linewidth}
        \centering
        \captionsetup{justification=centering}
        \begin{subfigure}{\linewidth}\centering
        \includegraphics[width=0.75\linewidth]{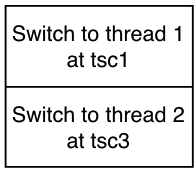}
        \caption{sideband trace}
        \label{fig@decode_b}
        \end{subfigure}%
    \end{minipage} \hspace{2em}
    \begin{minipage}{0.22\linewidth}
        \centering
        \captionsetup{justification=centering}
        \begin{subfigure}{1\linewidth}
        \includegraphics[width=\linewidth]{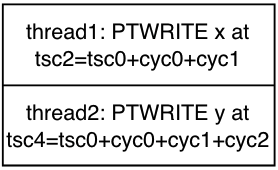}
        \caption{decoded events}
        \label{fig@decode_c}
        \end{subfigure}%
        \vspace{2em}
    \end{minipage} 
\caption{A trace example}
\label{fig@decode}
\end{figure}

The trace data is stored separately for each CPU. We first consider the case of a single CPU. For simplicity, we represent the PTWRITE trace packets as shown in \Cref{fig@decode_a}, which includes TSC, CYC, and PTWRITE packets. The TSC packet contains a specific timestamp. Similarly, each CYC packet indicates the number of clock cycles elapsed since the previous CYC or TSC packet. 
Based on these TSC and CYC packets, we can compute the timestamp for each PTWRITE packet. For example, the timestamp of \emph{<PTWRITE x>} in \Cref{fig@decode_a} can be calculated as $tsc2=tsc0 + cyc0 + cyc1$, while the timestamp of \emph{<PTWRITE y>} is $tsc4=tsc0 + cyc0 + cyc1 + cyc2$.
Meanwhile, the runtime data also includes a sideband trace obtained using perf events, which contains thread switch information with timestamps, shown as \Cref{fig@decode_b}. 
Suppose $tsc1<tsc2<tsc3<tsc4$, thus \emph{<PTWRITE x>} is executed after $tsc1$ but before $tsc3$. Therefore, we can derive that \emph{<PTWRITE x>} belongs to thread 1. Similarly, \emph{<PTWRITE y>} can be determined to belong to thread 2. 
As such, we obtain the final memory access events as shown in \Cref{fig@decode_c}. 
For multiple CPU cores, the event sequence from each CPU can be integrated according to their timestamps, resulting in a complete sequence of accesses.

At this stage, we essentially know the thread ID (tid) and the timestamp corresponding to each PTWRITE. Combined with the mapping information between each PTWRITE and the original instruction, we can determine the specific type of memory event (\eg, read, write, lock, or unlock). This ultimately allows us to reconstruct a trace similar to that shown in \Cref{fig@overview@exampled}.



\subsection{FastTrack-based Offline Detector}
\label{subsec:detect}

With the per-thread memory access events and thread synchronization events decoded in the previous stage, we can fully leverage dynamic data race detection algorithms for offline race detection. 
For the sake of fair comparison, \tool adopts the combination of happens-before and lockset algorithms which are utilized by the state-of-the-art dynamic race detectors \cite{tsan, fasttrack, prorace, kard}.
In brief, the offline detector reads and processes memory access events and thread synchronization events in the recorded order. It simulates dynamic data race detection by using information such as the recorded thread ID (TID), memory access addresses, lock variable addresses, and event types.
Note that again the contribution of this paper is to propose a low-overhead and high-precision data access monitoring approach for multithreaded programs via hardware tracing, which is orthogonal to the race detection algorithm. The race detection algorithms can benefit from our lightweight monitoring; and \tool can also adopt any other detection algorithms in the offline analysis. 

\section{Implementation}
\label{sec:implement}

In the static selective instrumentation module, we use Capstone \cite{capstone} and Angr \cite{angr} to construct an inter-procedural control flow graph (\ie, ICFG), and then perform our inter-procedural value set analysis and must race-free analysis over it. The instrumentation part is implemented based on Dyninst \cite{dyninst}, a well-known binary instrumentation framework.
The runtime trace collection module mainly utilizes the \emph{perf\_event\_open} function to configure the CPU buffer, and controls the relevant registers to enable Intel PTWRITE and IP filtering. By default, we allocate a 128MB memory buffer to maintain the hardware traces for each CPU core.
At the offline trace analysis part, the decoder is further developed based on the source code of the libipt library \cite{libipt} provided by Intel. 
We directly reuse the implementation of FastTrack-based detector in ProRace \cite{prorace} as our detection module.

\section{Evaluations}
\label{sec:evaluate}
In this section, we evaluate \tool over a comprehensive set of benchmarks to answer the following research questions. 
\begin{itemize}
    \item How well does \tool perform in terms of runtime overhead, static analysis scalability, instrumentation cost, and offline analysis efficiency? And how about comparison with the state-of-the-arts with respect to runtime overhead? (\S \ref{subsec:performance})
    \item What about the detection capability of \tool compared with the state-of-the-arts? (\S \ref{subsec:capability})
    \item How effective is our static selective instrumentation mechanism in reducing the runtime overhead and data loss? (\S \ref{subsec:effectiveness})
\end{itemize}

\begin{table*}[htbp]
\caption{The performance of \tool in terms of the time of static analysis (\S \ref{subsec:VSA}, \S\ref{subsec:MustRaceFree} and \S\ref{subsec:recordElim}), static binary instrumentation (\S \ref{subsec:SBI}), offline decoding (\S\ref{subsec:decode}), and offline race detection (\S \ref{subsec:detect}). \#Inst and \#Func represent the number of instructions and functions in the program, respectively. \label{table:overperformance}}
\scalebox{0.95}{
\begin{tabular}{lrrrrrr}
\toprule
subject & \#Inst & \#Func & static analysis & instrument & decoding  & race detection \\
\midrule
streamcluster & 4278 & 95 & 2.1s & 2.3s & 192.3s & 255.8s \\
x264 & 178647 & 1173 & 122.2s & 10.8s & 70.9s & 73.0s \\
vips & 815185 & 7333 & 462.3s & 59.3s & 32.2s & 20.8s \\
bodytrack & 107186 & 3417 & 32.7s & 3.2s & 43.7s & 47.1s \\
fluidanimate & 7220 & 108 & 2.8s & 2.5s & 117.6s & 122.0s \\
ocean\_cp & 24601 & 60 & 12.5s & 2.7s & 0.2s & 0.0s \\
ocean\_ncp & 15376 & 50 & 6.9s & 2.6s & 0.2s & 0.0s \\
raytrace & 19947 & 191 & 7.1s & 2.6s & 109.4s & 107.4s \\
water\_nsquared & 7754 & 61 & 3.3s & 2.4s & 0.7s & 0.6s \\
water\_spatial & 8152 & 62 & 3.3s & 2.4s & 0.2s & 0.0s \\
radix & 3314 & 48 & 2.2s & 2.4s & 0.8s & 0.5s \\
lu\_ncb & 3098 & 56 & 1.9s & 2.3s & 0.2s & 0.0s \\
lu\_cb & 3725 & 58 & 2.1s & 2.4s & 0.1s & 0.0s \\
barnes & 8385 & 107 & 3.5s & 2.4s & 114.7s & 131.9s \\
fft & 3908 & 57 & 2.2s & 2.5s & 0.1s & 0.0s \\
\midrule
(Arithmetic Mean) & 80718.4 & 858.4 & 44.5s & 6.9s & 45.6s & 50.6s \\
\bottomrule
\end{tabular}
}
\end{table*}

\subsection{Experimental Setup}
\MyPara{Benchmarks.} We primarily choose two representative sets of subject programs as our evaluation benchmarks. 
At first, to understand the performance of \tool, we take PARSEC/SPLASH-2x benchmark suite (version 3.0beta-20150206), which is commonly used to measure the performance in related works, such as ProRace \cite{prorace} and Kard \cite{kard}. The first three columns in Table \ref{table:overperformance} list the detailed characteristics about each subject program in terms of subject name, the number of instructions, the number of functions.

Moreover, to compare with the state-of-the-arts with respect to detection capability, we borrow the benchmark from ProRace \cite{prorace}, which is a set of real-world applications with known data races \cite{pset}. We follow ProRace’s measurement and test 11 buggy program versions, each with one different race triggered, including \emph{apache}, \emph{mysql}, \emph{cherokee}, \emph{pbzip} and \emph{aget}. Note that ProRace originally tested 12 program versions. However, since \emph{pfscan} is not publicly available in the repository, we can only test 11 of them.
Table \ref{table:detectioncapability} shows all the buggy versions and their manifest information. 

\MyPara{Comparison Tools.}
We select Kard \cite{kard}, ProRace \cite{prorace} and naive hardware tracing approach as the comparison approaches.
Kard leverages Intel MPK to allocate keys for inter-thread memory access protection, thus achieving low runtime overhead. 
ProRace, on the other hand, samples memory accesses using hardware PMU (in particular, Intel's Precise Event Based Sampling).
We select these two as comparison since they are the most recent work targeting low-overhead dynamic date race detection.
The naive hardware tracing approach is a variant of \tool by disabling the static selective instrumentation module. In other words, we naively treat all the registers involving shared memory accesses as the potential trace-points, and instrument PTWRITE instructions to record all of them at runtime.  

\MyPara{Environments.} We conduct all the experiments on a workstation with an Intel Core i9-14900K processor with 32 logical cores supporting Intel PTWRITE. 
The workstation has 64 GB memory and is equipped with a solid-state drive (SSD). 
It runs Ubuntu 22.04 LTS with a kernel version of 5.15.

\subsection{Overall Performance}
\label{subsec:performance}

We run the subjects five times and report the average time for each metric. Table \ref{table:overperformance} gives the detailed performance with respect to the time of static analysis (\ie, value-set analysis in \S \ref{subsec:VSA}, must race-free analysis in \S\ref{subsec:MustRaceFree} and redundant register elimination in \S\ref{subsec:recordElim}), static binary instrumentation (\S \ref{subsec:SBI}), offline trace decoding (\S\ref{subsec:decode}), and offline race detection (\S \ref{subsec:detect}).

As can be seen, the series of static selection analysis are efficient enough, which can be done with a couple of seconds for most subjects. The average time cost is only about 14 seconds. This is reasonable since we treat the shared value-set results in our analysis as flow-insensitive so as to achieve soundness and efficiency, which is elaborated in \S\ref{subsec:VSA}.
Moreover, the static binary instrumentation is also efficient enough as its complexity is linear to the program size.
For the offline part, both trace decoding and race detection can be easily finished with seconds on average. 
All of these validate that \tool is efficient enough, and can scale well in practice.

\begin{table*}[htbp]
\caption{The runtime overhead (TO) of \tool, compared with Kard, ProRace and Naive hardware tracing approach. The three columns under ProRace indicate the overhead with different sampling rate. $S_{inst}$ and $D_{inst}$ represent the number of instructions to be instrumented statically and to be traced dynamically, respectively.}
\label{table:overhead}
\begin{center}
\begin{adjustbox}{max width=\textwidth}
\begin{tabular}{lcccccccccc}
\toprule
\multirow{2}{*}{Bench Name} & \multicolumn{3}{c}{HardRace} & \multicolumn{1}{c}{Kard} & \multicolumn{3}{c}{ProRace (TO)} & \multicolumn{3}{c}{Naive}\\
\cmidrule(lr){2-4}
\cmidrule(lr){5-5}
\cmidrule(lr){6-8}
\cmidrule(lr){9-11}
 & TO & $S_{inst}$ & $D_{inst}$ & TO & 1/100 & 1/1000 & 1/10000 & TO & $S_{inst}$ & $D_{inst}$\\
\midrule
streamcluster & 0.2\% & 125 & 173.7M & 0.3\% & >30\% & >5\% & >5\% & 8.6\% & 487 & 464.2M \\
x264 & 3.5\% & 5994 & 52.1M & 3.0\% & - & - & - & 65.2\% & 20665 & 319.7M \\
vips & 0.1\% & 29426 & 23.8M & 1.3\% & - & - & - & 34.7\% & 99703 & 70.4M \\
bodytrack & 1.1\% & 234 & 37.0M & 10.4\% & >350\% & >5\% & >1\% & 36.3\% & 5855 & 0.0 \\
fluidanimate & 9.6\% & 73 & 96.5M & 61.9\% & >600\% & >90\% & >5\% & 24.3\% & 620 & 104.1M \\
ocean\_cp & 4.9\% & 132 & 6.7K & -5.9\% & - & - & - & 10.4\% & 4266 & 151.2M \\
ocean\_ncp & 4.0\% & 70 & 6.5K & 0.0\% & - & - & - & 57.8\% & 2247 & 271.5M \\
raytrace & 4.3\% & 444 & 89.5M & 3.7\% & >290\% & >30\% & >1\% & 30.7\% & 1726 & 57.3M \\
water\_nsquared & 0.3\% & 118 & 320.7K & 18.0\% & - & - & - & 4.1\% & 685 & 255.9M \\
water\_spatial & 0.2\% & 123 & 21.5K & 5.6\% & - & - & - & 3.7\% & 799 & 92.4M \\
radix & -0.0\% & 129 & 356.9K & -1.0\% & - & - & - & 7.7\% & 531 & 141.1M \\
lu\_ncb & -8.1\% & 91 & 6.5K & -5.2\% & - & - & - & 5.3\% & 393 & 27.8M \\
lu\_cb & -2.1\% & 94 & 6.5K & -4.7\% & - & - & - & 9.3\% & 531 & 33.9M \\
barnes & 3.6\% & 229 & 96.0M & 34.1\% & - & - & - & -5.7\% & 827 & 192.2M \\
fft & 2.5\% & 115 & 201.0 & 1.0\% & - & - & - & 4.6\% & 556 & 21.1M \\
\midrule
(Arithmetic Mean) & 1.6\% & 2493.1 & 38.0M & 8.2\% & >317.5\% & >32.5\% & >3.0\% & 19.8\% & 9326.1 & 146.9M \\
\bottomrule
\end{tabular}
\end{adjustbox}
\end{center}
\end{table*}

In addition, we also compare the runtime overhead of \tool with the state-of-the-arts, Kard \cite{kard}, ProRace \cite{prorace}, and the naive hardware tracing approach. \Cref{table:overhead} shows the detailed results. 
ProRace is a sampling-based approach, whose overhead relies on the exact sampling rate. As such, we measure the overheads of ProRace under three different sampling rates (\ie, 1/100, 1/1000, 1/10000). 
To calculate the overhead, we run the baseline subjects and the instrumented subjects each five times and compute the average. The overhead is computed as the execution time of instrumented program divided by the execution time of original program minus 100\%.
Note that due to nondeterministic execution of the subjects and the tiny execution time, it is possible that the overhead is negative.  
To understand further why the overheads differ greatly between \tool and naive tracing, we calculate the number of PTWRITE instructions to be instrumented statically and to be traced dynamically, denoted as $S_{inst}$ and $D_{inst}$ in Table \ref{table:overhead}, respectively.
Furthermore, the subjects evaluated by ProRace and Kard are not totally identical. We chose to align them with Kard since it is more recent then ProRace. The symbol ``-'' of Table \ref{table:overhead} indicates that the subject is not evaluated and no data is available for ProRace. 


We can observe that \tool achieves a negligible overhead of 1.6\% on average. Kard suffers from an overhead of 8.2\%. But importantly, it only supports a limited types of races. We will discuss them shortly in \S \ref{subsec:capability}.  
For ProRace, the overheads under different sampling rates vary significantly. It reaches beyond 300\% with 1/100 sampling rate. It also has around 3\% overhead under 1/10000. Again as is well known, the low sampling rate usually corresponds to low detection capability (see Table \ref{table:detectioncapability}). 
In contrast, \tool performs selective tracing, which achieves low overhead while not sacrificing detection capability.
Additionally, the average overhead of naive hardware tracing approach is 19.8\%. Even worse, an immense amount of data loss happens -- nearly 40\% of traces are lost shown as Table \ref{table:loss}.
We also compared the number of static instrumentation points ($S_{inst}$) and dynamically decoded instrumentation points ($D_{inst}$) between HardRace and the naive method. 
To understand the overhead differences between \tool and Naive hardware tracing, the data of $S_{inst}$ and $D_{inst}$ provides clues. 
The numbers of PTWRITE instructions to be instrumented statically (\ie,  $S_{inst}$) and to be executed dynamically (\ie,  $D_{inst}$) by \tool are about 2400+ and 38M, which are dramatically smaller than that of naive hardware tracing (\ie, 9300+ and 146M), respectively. This also indicates that our static selective instrumentation module significant prunes away unnecessary trace points, thus reducing the overhead. 


\begin{table*}[htbp]
\caption{Detection probability for each approach.}
\label{table:detectioncapability}
\begin{center}
\resizebox{\textwidth}{!}{
\begin{tabular}{lllccccc}
\toprule
 & \multirow{2}{*}{Bug manifestation} & \multirow{2}{*}{Type} & \multicolumn{3}{c}{ProRace/\%} & \multirow{2}{*}{HardRace/\%} & \multirow{2}{*}{Kard/\%}\\
\cmidrule(lr){4-6}
 &  &  & 1/100 & 1/1000 & 1/10000 &  \\
\midrule
apache-21287        & double free         & non-ILU   & 50         & 3                             & 0            & 100    & 0  \\
apache-25520        & corrupted log       & non-ILU & 57         & 52                            & 15           & 100    & 0  \\
apache-45605        & assertion           & non-ILU & 60         & 11                            & 1            & 100    & 0  \\
mysql-3596          & crash               & ILU   & 5          & 1                             & 0            & 100    & 100  \\
mysql-644           & crash               & ILU   & 21         & 6                             & 1            & 100    & 100  \\
mysql-791           & missing output      & ILU   & 59         & 2                             & 0            & 100    & 100  \\
cherokee-0.9.2      & corrupted log       & non-ILU & 63         & 29                            & 8            & 100    & 0  \\
cherokee-bug1       & corrupted log       & non-ILU & 57         & 19                            & 5            & 100    & 0  \\
pbzip2-0.9.4-crash  & crash               & ILU   & 0          & 0                             & 0            & 100    & 100  \\
pbzip2-0.9.4-benign & -                   & ILU       & 100        & 100                           & 100          & 100    & 100  \\
aget-bug2           & wrong record in log & ILU       & 100        & 100                           & 100          & 100    & 100  \\
\midrule
(Arithmetic Mean)                    &                     &          & 52.0         & 29.4       &  20.9            & 100.0    & 54.5\\
\bottomrule
\end{tabular}
}
\end{center}
\end{table*}

\subsection{Detection Capability}
\label{subsec:capability}
In this section, we measure the detection capability of \tool, and compare it with the state-of-the-arts. 
Each subject presented in \Cref{table:detectioncapability} is reported to contain a hard-to-trigger data race. And the reporters provided corresponding patches to control thread interleaving, enabling the data race to be triggered within a short period of time. 
We evaluate the detection capability in terms of detection probability which is introduced by ProRace \cite{prorace}.
To be specific, we run each buggy program 100 times and to count how many runs each detection tool can report the race. A probability of 50\% indicates that, among 100 runs, the data race is detected in 50 of them. 
A probability of 100\% means that the tool can detect the race every time, indicating that there are no false negatives for that subject.

Table \ref{table:detectioncapability} lists the detection probability data of various tools. 
\tool is able to detect the bugs in all the runs without any false negatives. 
This is because \tool only prunes away unnecessary memory accesses in a safe manner. 
Technically, the memory accesses that can trigger data races should all be recorded in hardware traces, allowing for detection when a data race occurs. 
In contrast, ProRace shows average detection probabilities of 52.0\%, 29.4\%, and 20.9\% for sampling rates 1/100, 1/1K, and 1/10K, respectively. This means that even with very dense sampling, the detection probability only reaches about a half, while the overhead is prohibitively high (as mentioned earlier in Table \ref{table:overhead}).
For Kard, due to the lack of relevant experimental data and the absence of source code, a direct comparison can hardly be conducted. 
However, based on our understanding of its approach, we can reason if each race can be detected by Kard. Specifically, Kard states that it can only  detect Inconsistent Lock Usage (ILU) races. 
Upon manual checking of the data race types of the subjects, we find that five of the eleven programs exhibit non-ILU type of data races.  Thus, they cannot be detected by Kard. 

\begin{table*}[htbp]
\caption{The data loss times and percentage due to buffer overflow}
\begin{center}
\label{table:loss}
\begin{tabular}{lcccc}
\toprule
\multirow{2}{*}{subject} & \multicolumn{2}{c}{\tool} & \multicolumn{2}{c}{Naive Hardware Tracing}\\
\cmidrule(lr){2-3}
\cmidrule(lr){4-5}
 & loss percent & loss times & loss percent & loss times \\
\midrule
streamcluster & 0.0\% & 0 & 8.7\% & 7 \\
x264 & 0.0\% & 0 & 84.0\% & 317 \\
vips & 0.0\% & 0 & 33.5\% & 5 \\
bodytrack & 0.0\% & 0 & 60.6\% & 36 \\
fluidanimate & 0.0\% & 0 & 22.0\% & 14 \\
ocean\_cp & 0.0\% & 0 & 0.0\% & 0 \\
ocean\_ncp & 0.0\% & 0 & 76.5\% & 80 \\
raytrace & 0.0\% & 0 & 61.3\% & 10 \\
water\_nsquared & 0.0\% & 0 & 59.6\% & 16 \\
water\_spatial & 0.0\% & 0 & 0.0\% & 0 \\
radix & 0.0\% & 0 & 51.6\% & 12 \\
lu\_ncb & 0.0\% & 0 & 13.1\% & 1 \\
lu\_cb & 0.0\% & 0 & 44.7\% & 1 \\
barnes & 0.0\% & 0 & 43.2\% & 17 \\
fft & 0.0\% & 0 & 0.0\% & 0 \\
\midrule
(Arithmetic Mean) & 0.0\% & 0 & 37.3\% & 34 \\
\bottomrule
\end{tabular}
\end{center}
\end{table*}

\subsection{Effectiveness of Selective Instrumentation}
\label{subsec:effectiveness}

As mentioned before, our static selective instrumentation module plays the crucial role in reducing runtime overhead and hardware data loss. 
In this section, we would like to validate the significance of selective instrumentation (\S \ref{sec:online-trace}) empirically. 
First, regarding overhead reduction, Table \ref{table:overhead} provides the experimental data about the runtime overhead, the number of PTWRITE instructions instrumented statically and traced dynamically by \tool and naive hardware tracing. Apparently, the overhead with selective instrumentation enabled (\ie, \tool) is much smaller than that of naive hardware tracing.  
Second, as for data loss, we measure the times of loss happened and the total percentage of data loss by \tool and naive hardware tracing. 
As shown in \Cref{table:loss}, \tool incurs no data loss for all subject programs under stress testing. 
In contrast, the naive hardware tracing has an average 37.3\% of data loss, with an average of 34 times of data loss events per subject. 
It is noteworthy that while the naive approach performs well on certain subjects such as \emph{streamcluster}, \emph{ocean\_cp}, and \emph{water\_spatial}, it still experiences significant data loss in the majority of cases. 
For \emph{x264} and \emph{ocean\_ncp}, it successfully collects less than 30\% of the data. 
The reason is that for the subjects like \emph{x264} and \emph{ocean\_ncp}, intensive memory accesses occur frequently in the program, leading to severe hardware data loss.  
All in all, we conclude that the selective analysis in \tool is highly effective in reducing data loss, thus ensuring detection capability.

\section{Related Work}
\label{sec:related}
Over the past years, various approaches for data race detection have been proposed, including static, dynamic, and hybrid approaches.

\MyPara{Dynamic Race Detection.}
The lockset algorithm, introduced by Eraser \cite{eraser}, checks whether shared memory is consistently protected by locks to detect data races. Although prone to false positives, its efficiency makes it a reference point for many subsequent studies. For example, the well-known ThreadSanitizer (TSAN) \cite{tsan} employs a combination of happens-before and lockset algorithms to balance detection accuracy and performance. Another prominent dynamic tool, FastTrack \cite{fasttrack}, optimizes the original happens-before approach for better performance. Other dynamic detectors like ProRace \cite{prorace}, Kard\cite{kard}, and TxRace \cite{txrace}, integrate hardware-based techniques into their detection processes. ProRace uses Intel Processor Trace (PT) to log the program's control flow and PEBS to sample memory accesses, then applies the FastTrack algorithm offline to detect races. Kard and TxRace, on the other hand, rely entirely on hardware-based methods. Kard uses Memory Protection Keys (MPK) to ensure that a shared object is accessible by only one thread within a critical section. TxRace utilizes hardware transactional memory (HTM) to detect data races dynamically, treating critical regions as atomic transactions and checking for conflicts. At the same time, some studies optimize the happens-before (HB) relationship at the algorithmic level to mitigate the shortcomings of dynamic data race detection. For instance, \cite{cp-relation} introduces the causally-precedes (CP) relationship, which is a subset of the HB relationship, allowing for the observation of more races without sacrificing robustness. Meanwhile, WCP\cite{wcp-relation} further weakens the CP relationship and enables race detection within linear time.

\MyPara{Static Race Detection.}
Static detectors often exhibit high false positives and low false negatives. For instance, RacerD \cite{racerd} is classified as a lockset-based detector, allowing it to avoid dealing with the vast number of interleavings typical in static analysis. Conversely, O2 \cite{origin}, which combines lockset and happens-before analysis, must construct a static happens-before graph (SHB) during static analysis and explore as many interleavings as possible to minimize false negatives.

\MyPara{Hybrid Race Detection.}
Although relatively uncommon, some work has explored the combination of static analysis followed by dynamic execution. For instance, RaceMob \cite{racemob} integrates both static and dynamic techniques: it first uses the static analysis tool RELAY \cite{relay} to detect potential data races, then breaks down these potential race conditions into tasks that are crowdsourced to users, ensuring minimal overhead for individual users.


\section{Conclusion}
\label{sec:conclude}
\tool is a data race monitor which can on-the-fly detect races happened in production runs. Its core technical contribution lies on  a series of sound static analysis which are unitized to prune away unnecessary memory accesses significantly, thus achieving super-low runtime overhead. To the best of our knowledge, \tool is the first work to leverage Intel PTWRITE for production-run data race detection. The experimental evaluations validate that \tool can achieve sufficiently low overhead while ensuring good detection capability. It to some extent proves that the holistic design combining static analysis and hardware tracing is promising for multithreaded program monitoring. We are looking forward to more attempts along this direction for multithreaded programs.

\section*{Data-Availability Statement}\label{sec:data-availability-statement}
We would like to provide the artifact later and submit it for Artifact Evaluation.
It would contain the source code of \tool, the benchmarks used, as well as all the experimental results.


\bibliographystyle{ACM-Reference-Format}
\bibliography{sample-base,mybibfile}


\begin{thebibliography}{32}


\ifx \showCODEN    \undefined \def \showCODEN     #1{\unskip}     \fi
\ifx \showDOI      \undefined \def \showDOI       #1{#1}\fi
\ifx \showISBNx    \undefined \def \showISBNx     #1{\unskip}     \fi
\ifx \showISBNxiii \undefined \def \showISBNxiii  #1{\unskip}     \fi
\ifx \showISSN     \undefined \def \showISSN      #1{\unskip}     \fi
\ifx \showLCCN     \undefined \def \showLCCN      #1{\unskip}     \fi
\ifx \shownote     \undefined \def \shownote      #1{#1}          \fi
\ifx \showarticletitle \undefined \def \showarticletitle #1{#1}   \fi
\ifx \showURL      \undefined \def \showURL       {\relax}        \fi
\providecommand\bibfield[2]{#2}
\providecommand\bibinfo[2]{#2}
\providecommand\natexlab[1]{#1}
\providecommand\showeprint[2][]{arXiv:#2}

\bibitem[Ahmad et~al\mbox{.}(2021)]%
        {kard}
\bibfield{author}{\bibinfo{person}{Adil Ahmad}, \bibinfo{person}{Sangho Lee},
  \bibinfo{person}{Pedro Fonseca}, {and} \bibinfo{person}{Byoungyoung Lee}.}
  \bibinfo{year}{2021}\natexlab{}.
\newblock \showarticletitle{Kard: lightweight data race detection with
  per-thread memory protection}. In \bibinfo{booktitle}{\emph{Proceedings of
  the 26th ACM International Conference on Architectural Support for
  Programming Languages and Operating Systems}} (Virtual, USA)
  \emph{(\bibinfo{series}{ASPLOS '21})}. \bibinfo{publisher}{Association for
  Computing Machinery}, \bibinfo{address}{New York, NY, USA},
  \bibinfo{pages}{647–660}.
\newblock
\showISBNx{9781450383172}
\urldef\tempurl%
\url{https://doi.org/10.1145/3445814.3446727}
\showDOI{\tempurl}


\bibitem[Anh(2014)]%
        {capstone}
\bibfield{author}{\bibinfo{person}{Quynh~Nguyen Anh}.}
  \bibinfo{year}{2014}\natexlab{}.
\newblock \showarticletitle{Capstone: Next generation disassembly framework}.
\newblock \bibinfo{journal}{\emph{Proceedings of the 2014 Black Hat USA, Black
  Hat USA}}  \bibinfo{volume}{14} (\bibinfo{year}{2014}).
\newblock


\bibitem[Balakrishnan and Reps(2010)]%
        {wysinwyx}
\bibfield{author}{\bibinfo{person}{Gogul Balakrishnan} {and}
  \bibinfo{person}{Thomas Reps}.} \bibinfo{year}{2010}\natexlab{}.
\newblock \showarticletitle{WYSINWYX: What you see is not what you eXecute}.
\newblock \bibinfo{journal}{\emph{ACM Trans. Program. Lang. Syst.}}
  \bibinfo{volume}{32}, \bibinfo{number}{6}, Article \bibinfo{articleno}{23}
  (\bibinfo{date}{aug} \bibinfo{year}{2010}), \bibinfo{numpages}{84}~pages.
\newblock
\showISSN{0164-0925}
\urldef\tempurl%
\url{https://doi.org/10.1145/1749608.1749612}
\showDOI{\tempurl}


\bibitem[Bernat and Miller(2011)]%
        {dyninst}
\bibfield{author}{\bibinfo{person}{Andrew~R. Bernat} {and}
  \bibinfo{person}{Barton~P. Miller}.} \bibinfo{year}{2011}\natexlab{}.
\newblock \showarticletitle{Anywhere, any-time binary instrumentation}. In
  \bibinfo{booktitle}{\emph{Proceedings of the 10th ACM SIGPLAN-SIGSOFT
  Workshop on Program Analysis for Software Tools}} (Szeged, Hungary)
  \emph{(\bibinfo{series}{PASTE '11})}. \bibinfo{publisher}{Association for
  Computing Machinery}, \bibinfo{address}{New York, NY, USA},
  \bibinfo{pages}{9–16}.
\newblock
\showISBNx{9781450308496}
\urldef\tempurl%
\url{https://doi.org/10.1145/2024569.2024572}
\showDOI{\tempurl}


\bibitem[Blackshear et~al\mbox{.}(2018)]%
        {racerd}
\bibfield{author}{\bibinfo{person}{Sam Blackshear}, \bibinfo{person}{Nikos
  Gorogiannis}, \bibinfo{person}{Peter~W. O'Hearn}, {and} \bibinfo{person}{Ilya
  Sergey}.} \bibinfo{year}{2018}\natexlab{}.
\newblock \showarticletitle{RacerD: compositional static race detection}.
\newblock \bibinfo{journal}{\emph{Proc. ACM Program. Lang.}}
  \bibinfo{volume}{2}, \bibinfo{number}{OOPSLA}, Article
  \bibinfo{articleno}{144} (\bibinfo{date}{oct} \bibinfo{year}{2018}),
  \bibinfo{numpages}{28}~pages.
\newblock
\urldef\tempurl%
\url{https://doi.org/10.1145/3276514}
\showDOI{\tempurl}


\bibitem[Bond et~al\mbox{.}(2010)]%
        {pacer}
\bibfield{author}{\bibinfo{person}{Michael~D. Bond},
  \bibinfo{person}{Katherine~E. Coons}, {and} \bibinfo{person}{Kathryn~S.
  McKinley}.} \bibinfo{year}{2010}\natexlab{}.
\newblock \showarticletitle{PACER: proportional detection of data races}. In
  \bibinfo{booktitle}{\emph{Proceedings of the 31st ACM SIGPLAN Conference on
  Programming Language Design and Implementation}} (Toronto, Ontario, Canada)
  \emph{(\bibinfo{series}{PLDI '10})}. \bibinfo{publisher}{Association for
  Computing Machinery}, \bibinfo{address}{New York, NY, USA},
  \bibinfo{pages}{255–268}.
\newblock
\showISBNx{9781450300193}
\urldef\tempurl%
\url{https://doi.org/10.1145/1806596.1806626}
\showDOI{\tempurl}


\bibitem[Chen et~al\mbox{.}(2021a)]%
        {herqules}
\bibfield{author}{\bibinfo{person}{Daming~D. Chen}, \bibinfo{person}{Wen~Shih
  Lim}, \bibinfo{person}{Mohammad Bakhshalipour}, \bibinfo{person}{Phillip~B.
  Gibbons}, \bibinfo{person}{James~C. Hoe}, {and} \bibinfo{person}{Bryan
  Parno}.} \bibinfo{year}{2021}\natexlab{a}.
\newblock \showarticletitle{HerQules: securing programs via hardware-enforced
  message queues}. In \bibinfo{booktitle}{\emph{Proceedings of the 26th ACM
  International Conference on Architectural Support for Programming Languages
  and Operating Systems}} (Virtual, USA) \emph{(\bibinfo{series}{ASPLOS '21})}.
  \bibinfo{publisher}{Association for Computing Machinery},
  \bibinfo{address}{New York, NY, USA}, \bibinfo{pages}{773–788}.
\newblock
\showISBNx{9781450383172}
\urldef\tempurl%
\url{https://doi.org/10.1145/3445814.3446736}
\showDOI{\tempurl}


\bibitem[Chen et~al\mbox{.}(2021b)]%
        {selectivetaint}
\bibfield{author}{\bibinfo{person}{Sanchuan Chen}, \bibinfo{person}{Zhiqiang
  Lin}, {and} \bibinfo{person}{Yinqian Zhang}.}
  \bibinfo{year}{2021}\natexlab{b}.
\newblock \showarticletitle{{SelectiveTaint}: Efficient Data Flow Tracking With
  Static Binary Rewriting}. In \bibinfo{booktitle}{\emph{30th USENIX Security
  Symposium (USENIX Security 21)}}. \bibinfo{publisher}{USENIX Association},
  \bibinfo{pages}{1665--1682}.
\newblock
\showISBNx{978-1-939133-24-3}
\urldef\tempurl%
\url{https://www.usenix.org/conference/usenixsecurity21/presentation/chen-sanchuan}
\showURL{%
\tempurl}


\bibitem[Corporation({[n.\,d.]})]%
        {libipt}
\bibfield{author}{\bibinfo{person}{Intel Corporation}.}
  \bibinfo{year}{[n.\,d.]}\natexlab{}.
\newblock \bibinfo{booktitle}{\emph{libipt: an Intel(R) Processor Trace decoder
  library}}.
\newblock
\urldef\tempurl%
\url{https://github.com/intel/libipt}
\showURL{%
\tempurl}
\newblock
\shownote{Accessed: 2024}.


\bibitem[Flanagan and Freund(2009)]%
        {fasttrack}
\bibfield{author}{\bibinfo{person}{Cormac Flanagan} {and}
  \bibinfo{person}{Stephen~N. Freund}.} \bibinfo{year}{2009}\natexlab{}.
\newblock \showarticletitle{FastTrack: efficient and precise dynamic race
  detection}. In \bibinfo{booktitle}{\emph{Proceedings of the 30th ACM SIGPLAN
  Conference on Programming Language Design and Implementation}} (Dublin,
  Ireland) \emph{(\bibinfo{series}{PLDI '09})}. \bibinfo{publisher}{Association
  for Computing Machinery}, \bibinfo{address}{New York, NY, USA},
  \bibinfo{pages}{121–133}.
\newblock
\showISBNx{9781605583921}
\urldef\tempurl%
\url{https://doi.org/10.1145/1542476.1542490}
\showDOI{\tempurl}


\bibitem[Holey et~al\mbox{.}(2013)]%
        {haccrg}
\bibfield{author}{\bibinfo{person}{Anup Holey}, \bibinfo{person}{Vineeth
  Mekkat}, {and} \bibinfo{person}{Antonia Zhai}.}
  \bibinfo{year}{2013}\natexlab{}.
\newblock \showarticletitle{HAccRG: Hardware-Accelerated Data Race Detection in
  GPUs}. In \bibinfo{booktitle}{\emph{Proceedings of the 2013 42nd
  International Conference on Parallel Processing}}
  \emph{(\bibinfo{series}{ICPP '13})}. \bibinfo{publisher}{IEEE Computer
  Society}, \bibinfo{address}{USA}, \bibinfo{pages}{60–69}.
\newblock
\showISBNx{9780769551173}
\urldef\tempurl%
\url{https://doi.org/10.1109/ICPP.2013.15}
\showDOI{\tempurl}


\bibitem[Huang et~al\mbox{.}(2013)]%
        {clap-jeff}
\bibfield{author}{\bibinfo{person}{Jeff Huang}, \bibinfo{person}{Charles
  Zhang}, {and} \bibinfo{person}{Julian Dolby}.}
  \bibinfo{year}{2013}\natexlab{}.
\newblock \showarticletitle{CLAP: recording local executions to reproduce
  concurrency failures}. In \bibinfo{booktitle}{\emph{Proceedings of the 34th
  ACM SIGPLAN Conference on Programming Language Design and Implementation}}
  (Seattle, Washington, USA) \emph{(\bibinfo{series}{PLDI '13})}.
  \bibinfo{publisher}{Association for Computing Machinery},
  \bibinfo{address}{New York, NY, USA}, \bibinfo{pages}{141–152}.
\newblock
\showISBNx{9781450320146}
\urldef\tempurl%
\url{https://doi.org/10.1145/2491956.2462167}
\showDOI{\tempurl}


\bibitem[Kasikci et~al\mbox{.}(2017)]%
        {Lazy_sosp17}
\bibfield{author}{\bibinfo{person}{Baris Kasikci}, \bibinfo{person}{Weidong
  Cui}, \bibinfo{person}{Xinyang Ge}, {and} \bibinfo{person}{Ben Niu}.}
  \bibinfo{year}{2017}\natexlab{}.
\newblock \showarticletitle{Lazy Diagnosis of In-Production Concurrency Bugs}.
  In \bibinfo{booktitle}{\emph{Proceedings of the 26th Symposium on Operating
  Systems Principles}} (Shanghai, China) \emph{(\bibinfo{series}{SOSP '17})}.
  \bibinfo{publisher}{Association for Computing Machinery},
  \bibinfo{address}{New York, NY, USA}, \bibinfo{pages}{582–598}.
\newblock
\showISBNx{9781450350853}
\urldef\tempurl%
\url{https://doi.org/10.1145/3132747.3132767}
\showDOI{\tempurl}


\bibitem[Kasikci et~al\mbox{.}(2013)]%
        {racemob}
\bibfield{author}{\bibinfo{person}{Baris Kasikci}, \bibinfo{person}{Cristian
  Zamfir}, {and} \bibinfo{person}{George Candea}.}
  \bibinfo{year}{2013}\natexlab{}.
\newblock \showarticletitle{RaceMob: crowdsourced data race detection}. In
  \bibinfo{booktitle}{\emph{Proceedings of the Twenty-Fourth ACM Symposium on
  Operating Systems Principles}} (Farminton, Pennsylvania)
  \emph{(\bibinfo{series}{SOSP '13})}. \bibinfo{publisher}{Association for
  Computing Machinery}, \bibinfo{address}{New York, NY, USA},
  \bibinfo{pages}{406–422}.
\newblock
\showISBNx{9781450323888}
\urldef\tempurl%
\url{https://doi.org/10.1145/2517349.2522736}
\showDOI{\tempurl}


\bibitem[Kini et~al\mbox{.}(2017)]%
        {wcp-relation}
\bibfield{author}{\bibinfo{person}{Dileep Kini}, \bibinfo{person}{Umang
  Mathur}, {and} \bibinfo{person}{Mahesh Viswanathan}.}
  \bibinfo{year}{2017}\natexlab{}.
\newblock \showarticletitle{Dynamic race prediction in linear time}. In
  \bibinfo{booktitle}{\emph{Proceedings of the 38th ACM SIGPLAN Conference on
  Programming Language Design and Implementation}} (Barcelona, Spain)
  \emph{(\bibinfo{series}{PLDI 2017})}. \bibinfo{publisher}{Association for
  Computing Machinery}, \bibinfo{address}{New York, NY, USA},
  \bibinfo{pages}{157–170}.
\newblock
\showISBNx{9781450349888}
\urldef\tempurl%
\url{https://doi.org/10.1145/3062341.3062374}
\showDOI{\tempurl}


\bibitem[Lamport(1978)]%
        {hb}
\bibfield{author}{\bibinfo{person}{Leslie Lamport}.}
  \bibinfo{year}{1978}\natexlab{}.
\newblock \showarticletitle{Time, clocks, and the ordering of events in a
  distributed system}.
\newblock \bibinfo{journal}{\emph{Commun. ACM}} \bibinfo{volume}{21},
  \bibinfo{number}{7} (\bibinfo{date}{July} \bibinfo{year}{1978}),
  \bibinfo{pages}{558–565}.
\newblock
\showISSN{0001-0782}
\urldef\tempurl%
\url{https://doi.org/10.1145/359545.359563}
\showDOI{\tempurl}


\bibitem[Lin et~al\mbox{.}(2019)]%
        {vsa-refine}
\bibfield{author}{\bibinfo{person}{Jian Lin}, \bibinfo{person}{Liehui Jiang},
  \bibinfo{person}{Yisen Wang}, {and} \bibinfo{person}{Weiyu Dong}.}
  \bibinfo{year}{2019}\natexlab{}.
\newblock \showarticletitle{A Value Set Analysis Refinement Approach Based on
  Conditional Merging and Lazy Constraint Solving}.
\newblock \bibinfo{journal}{\emph{IEEE Access}}  \bibinfo{volume}{7}
  (\bibinfo{year}{2019}), \bibinfo{pages}{114593--114606}.
\newblock
\urldef\tempurl%
\url{https://doi.org/10.1109/ACCESS.2019.2936139}
\showDOI{\tempurl}


\bibitem[Liu et~al\mbox{.}(2021)]%
        {origin}
\bibfield{author}{\bibinfo{person}{Bozhen Liu}, \bibinfo{person}{Peiming Liu},
  \bibinfo{person}{Yanze Li}, \bibinfo{person}{Chia-Che Tsai},
  \bibinfo{person}{Dilma Da~Silva}, {and} \bibinfo{person}{Jeff Huang}.}
  \bibinfo{year}{2021}\natexlab{}.
\newblock \showarticletitle{When threads meet events: efficient and precise
  static race detection with origins}. In \bibinfo{booktitle}{\emph{Proceedings
  of the 42nd ACM SIGPLAN International Conference on Programming Language
  Design and Implementation}} (Virtual, Canada) \emph{(\bibinfo{series}{PLDI
  2021})}. \bibinfo{publisher}{Association for Computing Machinery},
  \bibinfo{address}{New York, NY, USA}, \bibinfo{pages}{725–739}.
\newblock
\showISBNx{9781450383912}
\urldef\tempurl%
\url{https://doi.org/10.1145/3453483.3454073}
\showDOI{\tempurl}


\bibitem[Lu et~al\mbox{.}(2008a)]%
        {study_asplos08}
\bibfield{author}{\bibinfo{person}{Shan Lu}, \bibinfo{person}{Soyeon Park},
  \bibinfo{person}{Eunsoo Seo}, {and} \bibinfo{person}{Yuanyuan Zhou}.}
  \bibinfo{year}{2008}\natexlab{a}.
\newblock \showarticletitle{Learning from mistakes: a comprehensive study on
  real world concurrency bug characteristics}.
\newblock \bibinfo{journal}{\emph{SIGOPS Oper. Syst. Rev.}}
  \bibinfo{volume}{42}, \bibinfo{number}{2} (\bibinfo{date}{mar}
  \bibinfo{year}{2008}), \bibinfo{pages}{329–339}.
\newblock
\showISSN{0163-5980}
\urldef\tempurl%
\url{https://doi.org/10.1145/1353535.1346323}
\showDOI{\tempurl}


\bibitem[Lu et~al\mbox{.}(2008b)]%
        {empirical-lu-cocurrency}
\bibfield{author}{\bibinfo{person}{Shan Lu}, \bibinfo{person}{Soyeon Park},
  \bibinfo{person}{Eunsoo Seo}, {and} \bibinfo{person}{Yuanyuan Zhou}.}
  \bibinfo{year}{2008}\natexlab{b}.
\newblock \showarticletitle{Learning from mistakes: a comprehensive study on
  real world concurrency bug characteristics}. In
  \bibinfo{booktitle}{\emph{Proceedings of the 13th International Conference on
  Architectural Support for Programming Languages and Operating Systems}}
  (Seattle, WA, USA) \emph{(\bibinfo{series}{ASPLOS XIII})}.
  \bibinfo{publisher}{Association for Computing Machinery},
  \bibinfo{address}{New York, NY, USA}, \bibinfo{pages}{329–339}.
\newblock
\showISBNx{9781595939586}
\urldef\tempurl%
\url{https://doi.org/10.1145/1346281.1346323}
\showDOI{\tempurl}


\bibitem[Musuvathi et~al\mbox{.}(2008)]%
        {chess}
\bibfield{author}{\bibinfo{person}{Madanlal Musuvathi}, \bibinfo{person}{Shaz
  Qadeer}, \bibinfo{person}{Thomas Ball}, \bibinfo{person}{Gerard Basler},
  \bibinfo{person}{Piramanayagam~Arumuga Nainar}, {and} \bibinfo{person}{Iulian
  Neamtiu}.} \bibinfo{year}{2008}\natexlab{}.
\newblock \showarticletitle{Finding and reproducing Heisenbugs in concurrent
  programs}. In \bibinfo{booktitle}{\emph{Proceedings of the 8th USENIX
  Conference on Operating Systems Design and Implementation}} (San Diego,
  California) \emph{(\bibinfo{series}{OSDI'08})}. \bibinfo{publisher}{USENIX
  Association}, \bibinfo{address}{USA}, \bibinfo{pages}{267–280}.
\newblock


\bibitem[O'Callahan and Choi(2003)]%
        {hybrid_ppopp03}
\bibfield{author}{\bibinfo{person}{Robert O'Callahan} {and}
  \bibinfo{person}{Jong-Deok Choi}.} \bibinfo{year}{2003}\natexlab{}.
\newblock \showarticletitle{Hybrid dynamic data race detection}. In
  \bibinfo{booktitle}{\emph{Proceedings of the Ninth ACM SIGPLAN Symposium on
  Principles and Practice of Parallel Programming}} (San Diego, California,
  USA) \emph{(\bibinfo{series}{PPoPP '03})}. \bibinfo{publisher}{Association
  for Computing Machinery}, \bibinfo{address}{New York, NY, USA},
  \bibinfo{pages}{167–178}.
\newblock
\showISBNx{1581135882}
\urldef\tempurl%
\url{https://doi.org/10.1145/781498.781528}
\showDOI{\tempurl}


\bibitem[Savage et~al\mbox{.}(1997)]%
        {eraser}
\bibfield{author}{\bibinfo{person}{Stefan Savage}, \bibinfo{person}{Michael
  Burrows}, \bibinfo{person}{Greg Nelson}, \bibinfo{person}{Patrick
  Sobalvarro}, {and} \bibinfo{person}{Thomas Anderson}.}
  \bibinfo{year}{1997}\natexlab{}.
\newblock \showarticletitle{Eraser: a dynamic data race detector for
  multithreaded programs}.
\newblock \bibinfo{journal}{\emph{ACM Trans. Comput. Syst.}}
  \bibinfo{volume}{15}, \bibinfo{number}{4} (\bibinfo{date}{nov}
  \bibinfo{year}{1997}), \bibinfo{pages}{391–411}.
\newblock
\showISSN{0734-2071}
\urldef\tempurl%
\url{https://doi.org/10.1145/265924.265927}
\showDOI{\tempurl}


\bibitem[Serebryany and Iskhodzhanov(2009)]%
        {tsan}
\bibfield{author}{\bibinfo{person}{Konstantin Serebryany} {and}
  \bibinfo{person}{Timur Iskhodzhanov}.} \bibinfo{year}{2009}\natexlab{}.
\newblock \showarticletitle{ThreadSanitizer: data race detection in practice}.
  In \bibinfo{booktitle}{\emph{Proceedings of the Workshop on Binary
  Instrumentation and Applications}} (New York, New York, USA)
  \emph{(\bibinfo{series}{WBIA '09})}. \bibinfo{publisher}{Association for
  Computing Machinery}, \bibinfo{address}{New York, NY, USA},
  \bibinfo{pages}{62–71}.
\newblock
\showISBNx{9781605587936}
\urldef\tempurl%
\url{https://doi.org/10.1145/1791194.1791203}
\showDOI{\tempurl}


\bibitem[Sheng et~al\mbox{.}(2011)]%
        {racez}
\bibfield{author}{\bibinfo{person}{Tianwei Sheng}, \bibinfo{person}{Neil
  Vachharajani}, \bibinfo{person}{Stephane Eranian}, \bibinfo{person}{Robert
  Hundt}, \bibinfo{person}{Wenguang Chen}, {and} \bibinfo{person}{Weimin
  Zheng}.} \bibinfo{year}{2011}\natexlab{}.
\newblock \showarticletitle{RACEZ: a lightweight and non-invasive race
  detection tool for production applications}. In
  \bibinfo{booktitle}{\emph{2011 33rd International Conference on Software
  Engineering (ICSE)}}. \bibinfo{pages}{401--410}.
\newblock
\urldef\tempurl%
\url{https://doi.org/10.1145/1985793.1985848}
\showDOI{\tempurl}


\bibitem[Shoshitaishvili et~al\mbox{.}(2016)]%
        {angr}
\bibfield{author}{\bibinfo{person}{Yan Shoshitaishvili}, \bibinfo{person}{Ruoyu
  Wang}, \bibinfo{person}{Christopher Salls}, \bibinfo{person}{Nick Stephens},
  \bibinfo{person}{Mario Polino}, \bibinfo{person}{Andrew Dutcher},
  \bibinfo{person}{John Grosen}, \bibinfo{person}{Siji Feng},
  \bibinfo{person}{Christophe Hauser}, \bibinfo{person}{Christopher Kruegel},
  {and} \bibinfo{person}{Giovanni Vigna}.} \bibinfo{year}{2016}\natexlab{}.
\newblock \showarticletitle{SOK: (State of) The Art of War: Offensive
  Techniques in Binary Analysis}. In \bibinfo{booktitle}{\emph{2016 IEEE
  Symposium on Security and Privacy (SP)}}. \bibinfo{pages}{138--157}.
\newblock
\urldef\tempurl%
\url{https://doi.org/10.1109/SP.2016.17}
\showDOI{\tempurl}


\bibitem[Smaragdakis et~al\mbox{.}(2012)]%
        {cp-relation}
\bibfield{author}{\bibinfo{person}{Yannis Smaragdakis}, \bibinfo{person}{Jacob
  Evans}, \bibinfo{person}{Caitlin Sadowski}, \bibinfo{person}{Jaeheon Yi},
  {and} \bibinfo{person}{Cormac Flanagan}.} \bibinfo{year}{2012}\natexlab{}.
\newblock \showarticletitle{Sound predictive race detection in polynomial
  time}. In \bibinfo{booktitle}{\emph{Proceedings of the 39th Annual ACM
  SIGPLAN-SIGACT Symposium on Principles of Programming Languages}}
  (Philadelphia, PA, USA) \emph{(\bibinfo{series}{POPL '12})}.
  \bibinfo{publisher}{Association for Computing Machinery},
  \bibinfo{address}{New York, NY, USA}, \bibinfo{pages}{387–400}.
\newblock
\showISBNx{9781450310833}
\urldef\tempurl%
\url{https://doi.org/10.1145/2103656.2103702}
\showDOI{\tempurl}


\bibitem[Voung et~al\mbox{.}(2007)]%
        {relay}
\bibfield{author}{\bibinfo{person}{Jan~Wen Voung}, \bibinfo{person}{Ranjit
  Jhala}, {and} \bibinfo{person}{Sorin Lerner}.}
  \bibinfo{year}{2007}\natexlab{}.
\newblock \showarticletitle{RELAY: static race detection on millions of lines
  of code}. In \bibinfo{booktitle}{\emph{Proceedings of the the 6th Joint
  Meeting of the European Software Engineering Conference and the ACM SIGSOFT
  Symposium on The Foundations of Software Engineering}} (Dubrovnik, Croatia)
  \emph{(\bibinfo{series}{ESEC-FSE '07})}. \bibinfo{publisher}{Association for
  Computing Machinery}, \bibinfo{address}{New York, NY, USA},
  \bibinfo{pages}{205–214}.
\newblock
\showISBNx{9781595938114}
\urldef\tempurl%
\url{https://doi.org/10.1145/1287624.1287654}
\showDOI{\tempurl}


\bibitem[Yu and Narayanasamy(2009)]%
        {pset}
\bibfield{author}{\bibinfo{person}{Jie Yu} {and} \bibinfo{person}{Satish
  Narayanasamy}.} \bibinfo{year}{2009}\natexlab{}.
\newblock \showarticletitle{A case for an interleaving constrained
  shared-memory multi-processor}. In \bibinfo{booktitle}{\emph{Proceedings of
  the 36th Annual International Symposium on Computer Architecture}} (Austin,
  TX, USA) \emph{(\bibinfo{series}{ISCA '09})}. \bibinfo{publisher}{Association
  for Computing Machinery}, \bibinfo{address}{New York, NY, USA},
  \bibinfo{pages}{325–336}.
\newblock
\showISBNx{9781605585260}
\urldef\tempurl%
\url{https://doi.org/10.1145/1555754.1555796}
\showDOI{\tempurl}


\bibitem[Zhang et~al\mbox{.}(2017)]%
        {prorace}
\bibfield{author}{\bibinfo{person}{Tong Zhang}, \bibinfo{person}{Changhee
  Jung}, {and} \bibinfo{person}{Dongyoon Lee}.}
  \bibinfo{year}{2017}\natexlab{}.
\newblock \showarticletitle{ProRace: Practical Data Race Detection for
  Production Use}. In \bibinfo{booktitle}{\emph{Proceedings of the
  Twenty-Second International Conference on Architectural Support for
  Programming Languages and Operating Systems}} (Xi'an, China)
  \emph{(\bibinfo{series}{ASPLOS '17})}. \bibinfo{publisher}{Association for
  Computing Machinery}, \bibinfo{address}{New York, NY, USA},
  \bibinfo{pages}{149–162}.
\newblock
\showISBNx{9781450344654}
\urldef\tempurl%
\url{https://doi.org/10.1145/3037697.3037708}
\showDOI{\tempurl}


\bibitem[Zhang et~al\mbox{.}(2016)]%
        {txrace}
\bibfield{author}{\bibinfo{person}{Tong Zhang}, \bibinfo{person}{Dongyoon Lee},
  {and} \bibinfo{person}{Changhee Jung}.} \bibinfo{year}{2016}\natexlab{}.
\newblock \showarticletitle{TxRace: Efficient Data Race Detection Using
  Commodity Hardware Transactional Memory}. In
  \bibinfo{booktitle}{\emph{Proceedings of the Twenty-First International
  Conference on Architectural Support for Programming Languages and Operating
  Systems}} (Atlanta, Georgia, USA) \emph{(\bibinfo{series}{ASPLOS '16})}.
  \bibinfo{publisher}{Association for Computing Machinery},
  \bibinfo{address}{New York, NY, USA}, \bibinfo{pages}{159–173}.
\newblock
\showISBNx{9781450340915}
\urldef\tempurl%
\url{https://doi.org/10.1145/2872362.2872384}
\showDOI{\tempurl}


\bibitem[Zuo et~al\mbox{.}(2021)]%
        {jportal}
\bibfield{author}{\bibinfo{person}{Zhiqiang Zuo}, \bibinfo{person}{Kai Ji},
  \bibinfo{person}{Yifei Wang}, \bibinfo{person}{Wei Tao},
  \bibinfo{person}{Linzhang Wang}, \bibinfo{person}{Xuandong Li}, {and}
  \bibinfo{person}{Guoqing~Harry Xu}.} \bibinfo{year}{2021}\natexlab{}.
\newblock \showarticletitle{JPortal: precise and efficient control-flow tracing
  for JVM programs with Intel processor trace}. In
  \bibinfo{booktitle}{\emph{Proceedings of the 42nd ACM SIGPLAN International
  Conference on Programming Language Design and Implementation}} (Virtual,
  Canada) \emph{(\bibinfo{series}{PLDI 2021})}. \bibinfo{publisher}{Association
  for Computing Machinery}, \bibinfo{address}{New York, NY, USA},
  \bibinfo{pages}{1080–1094}.
\newblock
\showISBNx{9781450383912}
\urldef\tempurl%
\url{https://doi.org/10.1145/3453483.3454096}
\showDOI{\tempurl}


\end{thebibliography}

\appendix
\end{document}